\documentclass[aps,twocolumn,showpacs,preprintnumbers]{revtex4-2}

\usepackage{graphicx} 
\usepackage{subfigure}
\usepackage{multirow}

\usepackage{longtable}
\usepackage[T1]{fontenc}
\usepackage{dcolumn}

\usepackage{bm}       
\usepackage{amsfonts}  
\usepackage{amsmath}  
\usepackage{amssymb}
\usepackage{xcolor}
\usepackage{verbatim}

\usepackage{hyperref}
\newcommand{\rev}[1]{\textcolor{black}{#1}} 
\newcommand{\revv}[1]{\textcolor{black}{#1}} 

\begin{document}

\title{Calibration-free single-frame super-resolution fluorescence microscopy}

\author{Anežka Dostálová}
\email{dostalova@optics.upol.cz}
\author{Dominik Vašinka}
\author{Robert Stárek}
\author{Miroslav Ježek}
\email{jezek@optics.upol.cz}
\affiliation{
Department of Optics, Faculty of Science, Palacký University Olomouc, 17. listopadu 12, 77900 Olomouc, Czechia}

\begin{abstract}  

Molecular fluorescence microscopy is a leading approach to super-resolution and nanoscale imaging in life and material sciences. However, super-resolution fluorescence microscopy is often bottlenecked by system-specific calibrations and long acquisitions of sparsely blinking molecules. We present a deep-learning approach that reconstructs super-resolved images directly from a single diffraction-limited camera frame. The model is trained exclusively on synthetic data encompassing a wide range of optical and sample parameters, enabling robust generalization across microscopes and experimental conditions. Applied to dense terrylene samples with 150~ms acquisition time, our method significantly reduces reconstruction error compared to Richardson-Lucy deconvolution, ThunderSTORM multi-emitter fitting\rev{, and the deep-learning-based method DECODE}. The results confirm the ability to resolve emitters separated by 35~nm at 580~nm wavelength, corresponding to a sevenfold resolution improvement beyond the Rayleigh criterion. \rev{Extensive tests demonstrate strong generalization ability of the developed model and its resilience across a broad range of noise levels, numerical apertures, and optical aberrations. We further apply the method to complex biological data, resolving densely packed chromatin structures.} By delivering unprecedented details from a single short camera exposure without any prior information and calibration, our approach enables plug-and-play super-resolution imaging of fast, dense, or light-sensitive samples on common wide-field microscopy setups.

\end{abstract}

\maketitle

\section{Introduction}

Optical imaging of single fluorescent molecules results in diffraction-limited point spread functions (PSFs). In modern biomedical imaging and microscopy, resolving structures at the molecular scale is essential, and therefore, many approaches seek to overcome this limit. The ability to resolve, localize, and count individual molecules is vital for understanding molecular processes, cellular dynamics, and material properties \cite{Dani2010, Gahlmann2014}.
Super-resolution imaging techniques have revolutionized biological and materials research, allowing the study of molecular structures and interactions with unprecedented detail. In cell biology, precise molecular localization is crucial for understanding subcellular architecture \cite{SubcellArch}, tracking protein dynamics \cite{ProteinDyn, FLAP}, and mapping molecular pathways involved in processes such as intracellular transport \cite{IntracellTransport}. In neuroscience, resolving individual synaptic proteins provides insights into synaptic plasticity and neurodegenerative diseases \cite{SynBi1, SynBi2}. In materials science and nanotechnology, the ability to detect and count single molecules aids in characterizing nanoscale heterogeneities and defects~\cite{Defects}.
When molecules are sparsely distributed and their PSFs are isolated, localization methods can enhance the resolution significantly, down to 10~nm \cite{Aquino2011, Huang2016}.
However, this becomes notably more challenging at higher molecule densities, as overlapping PSFs hinder accurate localization. 
This challenge can be addressed by two main approaches, image reconstruction and stochastic molecule localization methods.

Image reconstruction methods are based on deconvolution using either direct inverse or statistical methods, such as maximum likelihood estimation. By computationally restoring high-frequency details lost during imaging, these techniques enhance resolution and are widely employed in super-resolution imaging \cite{RM2, RM3, RM4, RM5, DeconSTORM}. 
In fluorescence microscopy, the observed image is formed as a convolution of the true fluorophore distribution with the PSF, along with added noise. Deconvolution seeks to recover the underlying distribution by estimating and reversing the blurring effects of the PSF. However, these methods rely on precise system-specific calibration and prior knowledge of the PSF. 
Their effectiveness diminishes with noise and may introduce artifacts, necessitating careful parameter tuning for each specific case.

Single-molecule localization microscopy (SMLM) offers an alternative solution by utilizing stochastically blinking emitters and accumulating data across many frames~\cite{SMLM}. By using photoswitchable fluorophores, SMLM techniques enable the precise localization of individual molecules in successive frames \cite{STORM, dSTORM, PALM, Testa2012, Lapkiewicz2025}. The final super-resolved image is composed of the localizations from individual sparse frames. This allows for molecule localization even in dense samples but requires extended acquisition times, which prevents the application of these methods to live or moving samples.
While SMLM techniques offer exceptional spatial resolution, they rely on the stochastic activation and deactivation of fluorescent emitters, and consequently require carefully optimized fluorophore selection, sample preparation, and precise control of the excitation \cite{fluortags, FM}. Their applicability is therefore restricted in delicate biological environments. Furthermore, localization needs thorough system-specific calibration and tuning to accurately extract molecule positions and intensities from noisy fluorescence images \cite{ThunderSTORM, Sage2015}. The number of needed sequence frames can be reduced by multi-emitter fitting algorithms \cite{MEF, DAOSTORM, WTM}. However, these methods are limited to certain molecule densities, struggle in the presence of high background noise, and are computationally demanding.

In recent years, integrating deep learning techniques into SMLM has significantly advanced the field, offering powerful tools for improving localization accuracy, accelerating image reconstruction, and addressing challenges associated with high emitter densities and low signal-to-noise ratios \cite{Manko2023, DLinSMLM1, DLinFI, Henriques2021}. An example of such a technique is DeepSTORM \cite{DeepSTORM}, a convolutional neural network (CNN)-based approach that accelerates the image reconstruction process. However, the training of the model needs to be fine-tuned for particular experimental conditions separately, and the method is limited to blinking datasets. Another example is localization-based DECODE \cite{DECODE}, a probabilistic model that predicts emitter positions and their corresponding uncertainties. As with all deep learning methods, DECODE performance decreases if the experimental conditions deviate from those in the training data. A step toward less calibration-dependent solution is the UniFMIR model \cite{UniFMIR}, a foundation model that focuses on image restoration from noisy data. It is pre-trained on diverse available datasets, making it more generalizable to various microscopy modalities and biological structures. Fine-tuning is still needed for different imaging tasks.
In general, achieving high-resolution imaging of individual molecules remains a fundamental challenge, particularly in dynamic or live samples, where time constraints preclude the acquisition of the thousands of frames required by SMLM techniques.

Here, we present a completely calibration-free deep learning model designed for super-resolving image reconstruction in samples with high emitter concentrations from a single frame, without any prior knowledge about the optical system, such as the shape of the point spread function or background noise distribution. Our model bypasses the drawbacks of localization-based methods by performing direct image reconstruction. \rev{It is a ready-to-use tool that requires only a single dense image as input, eliminating the need for specialized datasets with stochastically blinking fluorophores and broadening applicability to a wide range of fluorescent markers and optical systems.} Consequently, while conventional methods typically rely on acquiring thousands of frames—which can take several minutes per sample—our approach enables rapid measurement times. This significant reduction in acquisition time minimizes phototoxicity and photobleaching, which is particularly essential for imaging sensitive biological samples or when using sensitive organic fluorescent dyes.

Unlike the majority of state-of-the-art localization and reconstruction methods, which require extensive prior calibration information or frequent retraining, our model operates effectively without such demand, substantially simplifying experimental workflows and making it particularly well-suited for imaging dynamic samples. Additionally, under realistic experimental conditions, parameters of the optical setup vary in time and are non-uniform over the field of view. Also without the need for retraining or recalibration, our device-independent model is capable of handling these field-dependent variations in parameters such as different shapes or sizes of the PSFs, caused, for example, by aberrations, non-uniform background distribution, etc., within the field of view, \rev{which limits the accuracy of system-specific methods that cannot generalize beyond the calibration data}. Our approach offers a fast, flexible solution for super-resolution imaging, enhancing accessibility and minimizing setup requirements across various microscopy platforms.

\section{Results and Discussion}

\begin{figure}[h!]              
    \centering
    \includegraphics[width=0.99\columnwidth]{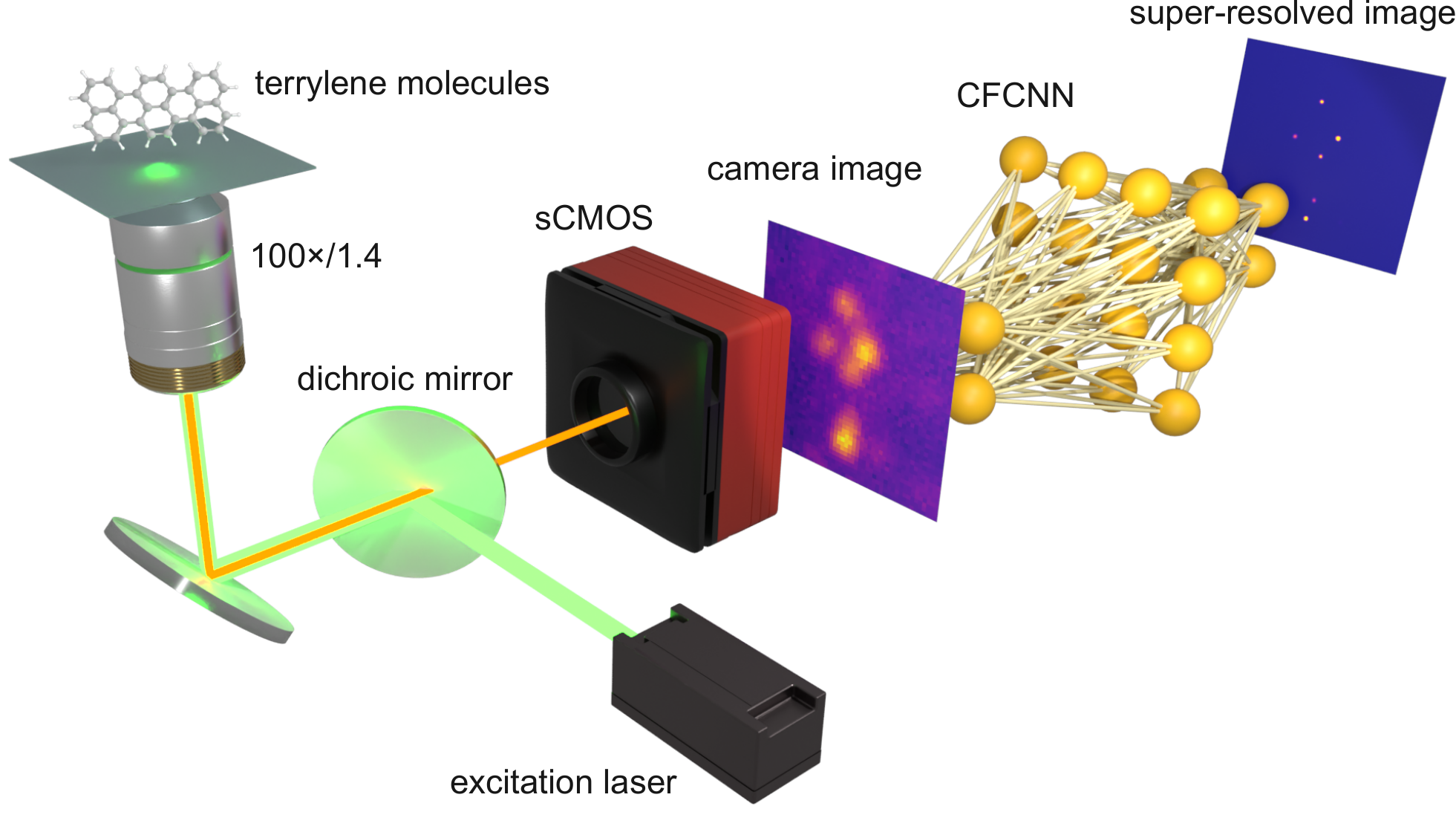}
    \caption{Schematic illustration of the experimental fluorescence microscope setup and the CFCNN application to the camera image. Terrylene molecules in the sample emit fluorescent light upon laser excitation. The fluorescence is collected by the microscope objective, separated from the excitation light on the dichroic mirror, and detected by an sCMOS camera. The resolution-limited camera image is directly processed by the CFCNN without requiring any prior information or calibration. The CFCNN reconstructs a super-resolved image, revealing individual molecules with significantly enhanced resolution.}
    \label{fig:expsetup}
\end{figure}

We applied our calibration-free convolutional neural network (CFCNN) to fluorescence intensity images of terrylene molecules at various concentrations embedded in a polyvinyl alcohol (PVA) layer. These images were acquired using a fluorescence microscope, schematically shown in Fig.~\ref{fig:expsetup}. The sample is illuminated by an excitation laser through a high numerical aperture oil-immersion objective lens, and the emitted fluorescence is collected by the objective and detected by an sCMOS camera. For details on the experiment, please refer to subsection A of the Methods. Due to the high molecule density and overlap of their PSFs, obtaining a super-resolved counterpart for the single dense camera image poses a significant challenge for traditional super-resolution methods.

\begin{figure}[h!]              
    \centering
    \includegraphics[width=0.99\columnwidth]{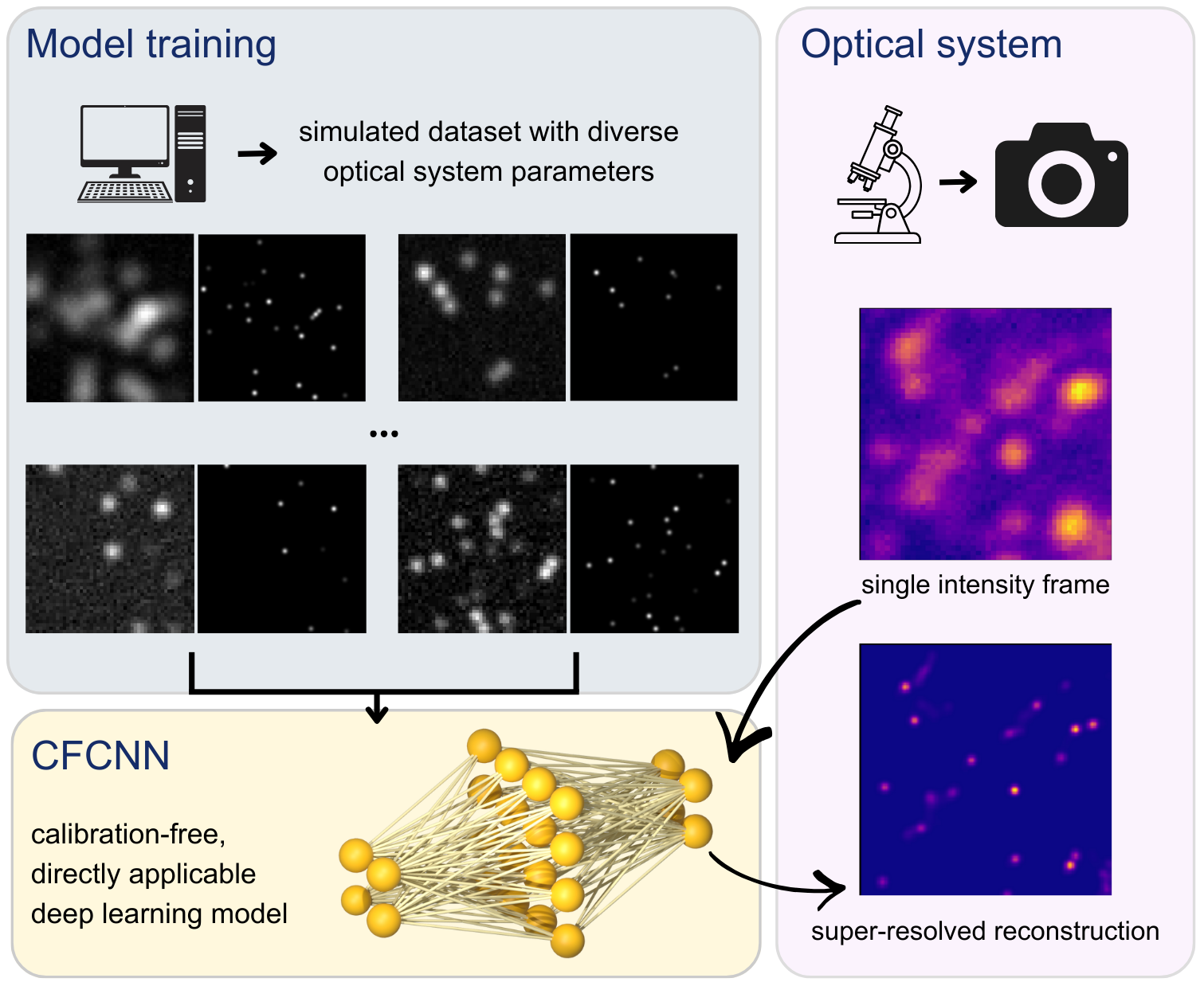}
    \caption{\rev{The calibration-free concept of the CFCNN. The training data are purely synthetic and cover wide ranges of optical system parameters, such as PSF shape and size, background noise, emitter concentration, etc. Pairs of ground truth and low-resolution image are generated for random combinations of the system parameters. The result is a calibration-free convolutional neural network, a ready-to-use tool that accepts a single intensity camera image at input and directly returns a super-resolved reconstruction, not requiring any other information or settings from the user. (The CFCNN depicted here is solely illustrative. For details on its architecture and technical parameters, please refer to the Methods section B.)}}
    \label{fig:cf-concept}
\end{figure}

\begin{figure*}[ht]              
    \centering
    \includegraphics[width=0.99\textwidth]{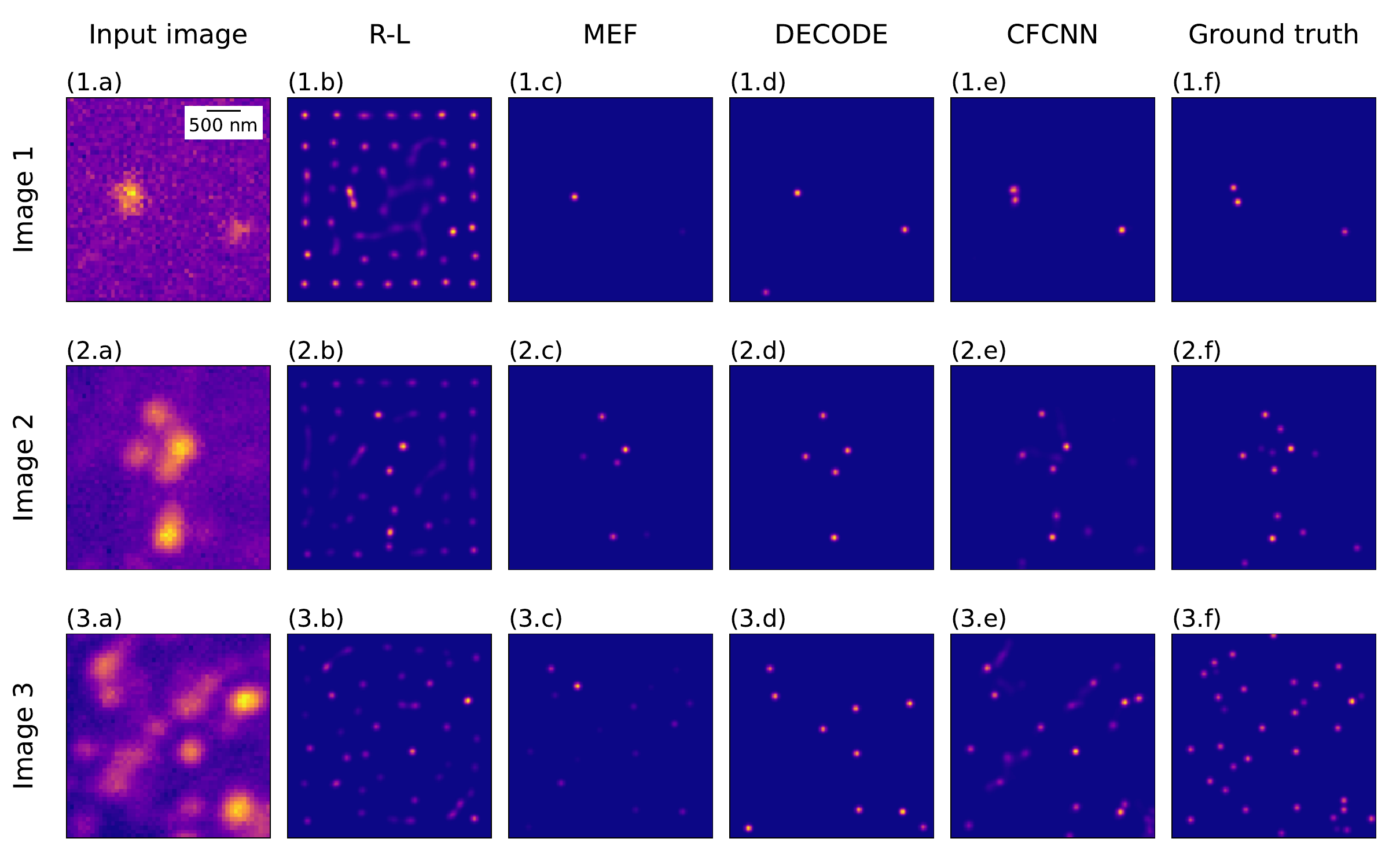}
    \caption{A visual comparison of the results from \rev{three} state-of-the-art methods for super-resolution imaging and localization, and our calibration-free neural network (CFCNN). All resulting images in panels (b)--(f) are convolved with a narrow Gaussian filter for better visualization. Panels (a) show the input intensity images of organic terrylene molecules from a fluorescence microscope acquired by an sCMOS camera, with different molecule powers and densities. 
    (b) Results of the reconstruction by the Richardson-Lucy deconvolution algorithm provided with a specific PSF. (c) The outputs of multi-emitter fitting using ThunderSTORM with specified camera settings. \rev{(d) Resulting images by DECODE trained on a full set of parameters of the imaging setup and sample.} (e) Super-resolved images reconstructed by our CFCNN using no prior information or calibration. 
    (f) An estimation of the ground truth using temporal information beyond the single intensity image.}
    \label{fig:tiPVA}
\end{figure*}

Following the experimental setup, the camera image is processed by our CFCNN model. This deep-learning approach is calibration-free due to its unique training methodology \cite{damn2026}, \rev{see Fig.~\ref{fig:cf-concept}}. The input to our model is a single intensity image, and the output is an upsampled, high-resolution reconstruction. The model is based on a convolutional neural network trained exclusively on synthetic data. 
This enables the model to learn across a vast range of optical system parameters, covering configurations that would be impossible to obtain experimentally at such a scale, and eliminating systematic errors inherent in experimental measurements.
Additionally, our network is not constrained to a single specific optical configuration. 
Instead, it is trained on diverse combinations of noise levels, PSF profiles, and other system parameters. This broad training scope enables the model to generalize effectively across various experimental conditions, making it robust both to variations in optical setup parameters across different systems and to variations within a single field of view, including aberrations, non-uniform background distributions, and other spatially varying changes. For further details on the CFCNN architecture and its training, please refer to the Methods, subsection B, \rev{and our code and data public repository \cite{zenodo}}.

% % Original table
% \renewcommand{\arraystretch}{1.75} 
% \begin{table*}[ht]
%     \centering
%     \begin{tabular}{p{2.6cm} p{2.8cm} >{\centering\arraybackslash}p{2.9cm} >{\centering\arraybackslash}p{2.9cm} >{\centering\arraybackslash}p{2.9cm} >{\centering\arraybackslash}p{2.9cm}}
%     \hline
%          & & R-L & MEF & DECODE & CFCNN \\\hline
%         Image 1 & MAE [$10^{-5}$] & 4.64(8) & 4.2*    & 3.0(5)  & 2.4(5) \\
%                 & KLD             & 6.0(2)  & 12.0*   & 14(5)   & 1(2)   \\
%         Image 2 & MAE [$10^{-5}$] & 3.98(7) & 3.5(1)  & 4.42(7) & 2.27(8)\\       
%                   & KLD           & 7.8(7)  & 18(2)   & 20(1)   & 2.1(4) \\
%         Image 3 & MAE [$10^{-5}$] & 3.52(3) & 3.97(2) & 4.50(6) & 3.34(4)\\
%                   & KLD           & 10.0(3) & 19.0(4) & 29.2(8) & 5.2(2) \\
%         \hline
%     \end{tabular}
%     \caption{Quantitative similarity metrics calculated between the super-resolved experimental fluorescence microscopy images relative to the ground truth for the Richardson-Lucy deconvolution, multi-emitter fitting using ThunderSTORM, DECODE, and the CFCNN. The digit in parentheses represents one standard deviation, e.g. 4.64(8) = 4.64$\pm$0.08. Asterisk denotes cases where the standard deviation is lower by at least 3 orders of magnitude than the measured value.}
%     \label{tab:metrics}
% \end{table*}
% \renewcommand{\arraystretch}{1} 

% newly trained DECODE (21/10/2025)
\renewcommand{\arraystretch}{1.75} 
\begin{table*}[ht]
    \centering
    \begin{tabular}{p{2.6cm} p{2.8cm} >{\centering\arraybackslash}p{2.9cm} >{\centering\arraybackslash}p{2.9cm} >{\centering\arraybackslash}p{2.9cm} >{\centering\arraybackslash}p{2.9cm}}
    \hline
         & & R-L& MEF & \rev{DECODE} & CFCNN \\\hline
        Image 1 & MAE [$10^{-5}$] & 4.64(8) & 4.2(3)    & \rev{3.6(5)}  & 2.4(5) \\
                & KLD             & 6(2)   & 12(2)     & \rev{14(5)}   & 1(2)   \\
        Image 2 & MAE [$10^{-5}$] & 3.98(7) & 3.5(1)    & \rev{4.49(7)} & 2.27(8)\\       
                & KLD             & 7.8(7)  & 18(1)     & \rev{22(1)}   & 2.1(4) \\
        Image 3 & MAE [$10^{-5}$] & 3.52(3) & 4.0(1)    & \rev{4.58(6)} & 3.34(4)\\
                & KLD             & 10.0(3) & 19.0(8)   & \rev{31.2(8)} & 5.2(2) \\
        \hline
    \end{tabular}
    \caption{Quantitative similarity metrics calculated between the super-resolved experimental fluorescence microscopy images relative to the ground truth for the Richardson-Lucy deconvolution, multi-emitter fitting using ThunderSTORM, \rev{DECODE}, and the CFCNN. The digit in parentheses represents one standard deviation, e.g. 4.64(8) = 4.64$\pm$0.08.}
    %Asterisk denotes cases where the standard deviation is lower by at least 3 orders of magnitude than the measured value.
    \label{tab:metrics}
\end{table*}
\renewcommand{\arraystretch}{1} 

To validate and benchmark the performance and versatility of our CFCNN, we also employed \rev{three} commonly used state-of-the-art \rev{and reference} methods on the same input images, specifically, the Richardson-Lucy (R-L) deconvolution algorithm as a reconstruction technique \cite{R-L_R, R-L_L}, multi-emitter fitting (MEF) with ThunderSTORM (an ImageJ plugin) as a localization method \cite{ThunderSTORM}, \rev{and DECODE as a deep learning approach \cite{DECODE}}. All approaches were provided exclusively with a single resolution-limited image. The amount of additional information and parameter settings required by each method differs, as well as the necessary post-processing steps to produce the final super-resolved image.
R-L algorithm iteratively applies Bayes theorem to update an estimate of the true image by maximizing the likelihood of the observed image. The algorithm can achieve significant resolution enhancement. However, it relies on accurate prior knowledge of the PSF profile, which necessitates either acquiring a calibration image of a sparse sample or the presence of an isolated molecule within the field of view. Besides being a time-consuming extra step, this also introduces a simplification, as the PSF shape may vary across the field of view.
ThunderSTORM offers extensive customization options tailored to the user’s specific application. However, this also means the results are highly dependent on a large number of input parameters, such as the camera settings, image filtering, fitting algorithm method, parameters, and so forth. Their values need to be optimized for each given setup and measurement. The optimization involves balancing multiple trade-offs, requiring user expertise and adding complexity to the process.
The output of the method is a table containing the emitter coordinates and intensities. Subsequent image rendering may lead to ambiguity since there are multiple ways to perform this step, and requires additional decisions in the processing pipeline. 
\rev{DECODE requires training on specific detailed calibration data, the acquisition of which is an experimentally demanding and time-consuming task. For the training, a fitted PSF $z$-stack needs to be provided, and parameters of the experiment must be specified, such as the camera parameters, emitter density, brightness, and so on. Once trained, it can be used only on data acquired under the same conditions, otherwise, retraining is necessary.}
Details on the settings for each method are given in part C of the Methods section.
\rev{In contrast, our deep-learning approach operates in a plug-and-play manner, requiring no additional training or information beyond the input image, neither before nor after applying the CFCNN. The resulting super-resolved image outperforms the other methods across all evaluated metrics.}

The visual comparison between the input intensity image of fluorescent terrylene molecules, the reconstructed outputs from R-L deconvolution, multi-emitter fitting using ThunderSTORM, \rev{DECODE}, and our CFCNN, is presented in Fig.~\ref{fig:tiPVA}. The output images are accompanied by the ground truth, obtained by leveraging the limited photostability of the molecules in PVA and their gradual localization from a recorded time-lapse of bleaching events (see the Methods section, part D, for details). 
We show the results for three independent experimental intensity images with different molecule concentrations and levels of background noise.
The input image size is $50 \times 50$ pixels and all outputs are upscaled to $200 \times 200$ pixels to enhance the resolution and detail of the reconstructed image.
The multi-emitter fitting, \rev{DECODE}, and our manual localization technique for obtaining the ground truth all produce a list of localized emitters and their intensities, which are then rendered into the finer grid. The situation is different for R-L deconvolution, which directly returns an image of the original size. In this case, the input image was first upscaled via bicubic interpolation before applying the reconstruction algorithm. Finally, the CFCNN has upsampling layers incorporated in its architecture.
Since the reconstructions consist of $200 \times 200$ pixels containing small structures, they are convolved with a narrow Gaussian filter (with $\sigma = 2$~px) for visualization purposes in the figure. 
For further details on the methods employed and the process used to obtain the ground truth, please refer to the Methods section.

The performance of R-L deconvolution, shown in panels (b), is highly sensitive to the noise level in the data, and the algorithm accuracy deteriorates as the SNR decreases.
It introduces artifacts by placing emitters at approximately equidistant positions, effectively filling noisy regions with emitters not present in the original data. This leads to a loss of accuracy and the appearance of artificial structures in the reconstructed image. This effect is most pronounced in Fig.~\ref{fig:tiPVA} (1.b) and is also clearly visible in panel (2.b).

\begin{figure*}[ht]              
    \centering
    \includegraphics[width=0.95\textwidth]{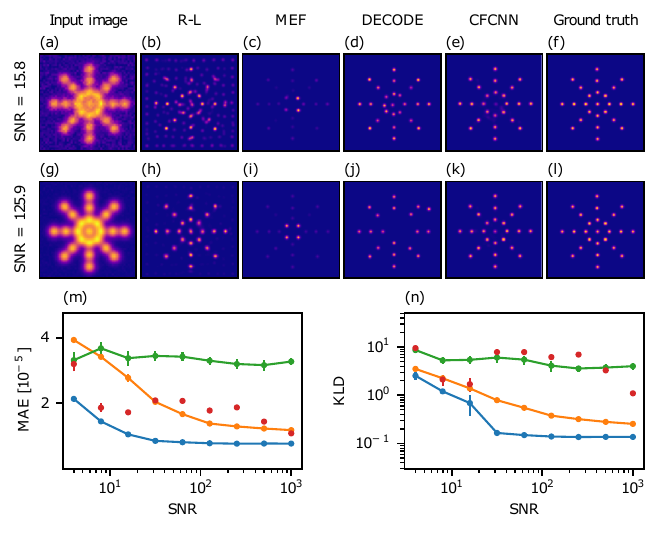}
    \caption{\rev{Synthetic data with emitters arranged in the shape of a star. Images in panels (b)--(f) and (h)--(l) are convolved with a narrow Gaussian filter for better visualization. (a, g) Examples of the input image with $\textrm{SNR}=15.8$ and $\textrm{SNR}=125.9$, respectively, obtained as the ground truth (f, l) convolved with a Gaussian PSF with the width of $2$~px on the $50 \times 50$ pixel grid and the corresponding noise backgrounds added. (b, h) The image reconstructions by the Richardson-Lucy deconvolution algorithm. (c, i) The outputs of multi-emitter fitting using ThunderSTORM. (d, j) The resulting images produced by DECODE. A uniquely trained model was used for each SNR. (e, k) The reconstructed images by the CFCNN without recalibration and retraining. (m) Mean absolute error between the outputs and the ground truth as a function of SNR for R-L (orange), MEF (green), DECODE (red), and the CFCNN (blue). (n) Kullback–Leibler divergence of the outputs from the ground truth as a function of SNR. In the case of DECODE, a separate model was trained on the corresponding SNR and PSF for each data point.}}
    \label{fig:synth_data}
\end{figure*}

The resulting images from the MEF regime of ThunderSTORM are shown in panels (c) of Fig.~\ref{fig:tiPVA}. 
Without temporal information to distinguish overlapping emitters, the algorithm tends to produce a single emitter at the center of mass of a diffraction-limited spot, as clearly seen in Fig.~\ref{fig:tiPVA} (1.c). It cannot determine the actual number of molecules, leading to localization errors. It is evident from all three images that MEF missed a considerable number of emitters or inaccurately estimated their intensities, and some were localized incorrectly.

\rev{For images produced by DECODE, see panels (d) of Fig.~\ref{fig:tiPVA}. Although DECODE was trained on the specific PSF and optical parameters compatible with the experimental images, the reconstructions are incomplete and close emitters are not resolved, as evident in Fig.~\ref{fig:tiPVA}~(1.d).}

In contrast, our CFCNN demonstrates superior performance across the field of view regardless of background noise level. It accurately reconstructs individual molecules directly from the single intensity frame, without any prior information about the input image origin. It surpasses the Rayleigh criterion limit and achieves a level of super-resolution that typically requires large datasets of sparsely distributed emitters. \rev{Panels (e)} in Fig.~\ref{fig:tiPVA} clearly illustrate the ability of the CFCNN to distinguish emitters even in dense regions of the sample.

For quantitative performance evaluation, we compared the output of each method against the ground truth utilizing the mean absolute error (MAE) and Kullback–Leibler divergence (KLD). MAE measures pixel-wise accuracy, while KLD captures the distance of probability distributions. Together, they provide a comprehensive assessment of both reconstruction precision and distributional similarity.
These metrics are evaluated relative to the ground truth of each sample image and summarized in Table~\ref{tab:metrics} with statistical uncertainties. Given the range of factors affecting the absolute values, we present three independent experimental images (shown in Fig.~\ref{fig:tiPVA}) to highlight the trends of the methods and provide a comparative analysis of their performance.
It is evident that the CFCNN consistently outperforms the other methods, even with no requirement of prior information and calibration. 
\revv{It may be surprising at first that a ready-to-use model with no input information yields better results than system-specific methods, carefully calibrated for a given scenario. In general, this is attributed to field-dependent variations of system and sample parameters. Specifically, the emitter concentration is typically not constant over the field of view (FOV), the PSF varies across the FOV due to inherent properties of the optical system, the background is not homogeneous, and therefore the SNR also varies. In the discussed experimental data, the dominant factors are the non-uniform emitter concentration and the SNR variations. State-of-the-art methods, particularly deep-learning approaches trained on specific parameter values (or calibration data), cannot generalize to this variability.}
This underscores the robustness and effectiveness of our approach, especially in challenging scenarios with high emitter concentrations, low SNR, and variable experimental conditions.

To further investigate the methods performance beyond the experimental results, we conducted additional tests on synthetic datasets with controlled parameters and precisely known ground truth, which allows for accurate and reliable evaluation. Additionally, it enables a systematic study of the performance dependence on the intensity noise level. Throughout this work, we define the SNR as the difference between the maximum intensity and the average noise level, divided by the standard deviation of background noise.
\rev{In Fig.~\ref{fig:synth_data} (a)--(l), we show a particular example of emitters arranged in the shape of a star. Panel~(a) represents the input for the tested methods with the $\textrm{SNR}=15.8$, while panel~(f) shows the generated ground truth with the applied narrow Gaussian filter for better visualization. For the outputs of each studied method, also convolved with the narrow Gaussian filter, see panels~(b)--(e). The second row showcases the input image with $\textrm{SNR}=125.9$ in panel~(g) and the corresponding results in (h)--(k). It is immediately apparent that the CFCNN reconstruction is virtually perfect, with all the emitters correctly positioned and their intensities accurately estimated, except for a slightly weaker rendering of the central emitter for $\textrm{SNR}=15.8$. On the contrary, the R-L reconstruction contains numerous artificial emitters, not only in the vicinity of the true emitter positions but across the entire field of view, particularly for the low SNR regime. The same also holds for the high SNR, but the artifacts are less bright. This renders the R-L results misleading and significantly complicates distinguishing genuine emitters from artifacts. The multi-emitter fitting struggles in dense regions and inaccurately estimates the emitter intensities, resulting in a compromised image. Some of the emitters were not resolved at all, leading to a loss of valuable information.
Finally, DECODE provides a reasonable output for $\textrm{SNR}=15.8$, however, with several emitters slightly misplaced and their intensities imprecisely estimated. For $\textrm{SNR}=125.9$ and other SNR values in the range approximately from 30 to 300, DECODE omits the entire central part of the star sample despite being trained for the specific SNR and PSF values. The main reason for this failure is the extremely low generalization ability of its trained models. As the density of emitters, their brightness, and SNR must be specified during the training stage, the resulting model works well only for images that match exactly those parameters. Any mismatch, even local, in a part of the field of view, results in significant errors in the DECODE output. These errors typically manifest as missing emitters, likely due to the relatively strong sparsity enforcement in DECODE training. Generally speaking, MEF and DECODE tend to omit emitters, which may represent tolerable errors in single-molecule localization microscopy, such as STORM and PALM, where the density of active molecules in a single frame is low and there is a high probability of repeated molecule localization in several frames. However, when super-resolution is performed from a single camera image, these methods fail to provide high-fidelity super-resolution imaging, especially for images with inhomogeneous emitter concentrations across the field of view.}

\rev{We also evaluated MAE and KLD between the outputs and the well-defined ground truth to numerically quantify the reconstruction performance. Fig.~\ref{fig:synth_data}~(m), (n) depicts the dependence of these metrics on the noise level expressed by SNR for all four methods. As before, R-L, MEF, and DECODE were supplied with all necessary information. Specifically, for DECODE, each data point corresponds to a unique model trained for the respective SNR value, requiring tedious recalibration and retraining that takes dozens of hours of extra work. Our CFCNN consistently outperforms the R-L algorithm, the MEF method, and DECODE across all SNR values, despite operating without any additional information, calibration, or retraining.}

\begin{figure}[h!]              
    \centering
    \includegraphics[width=0.8\columnwidth]{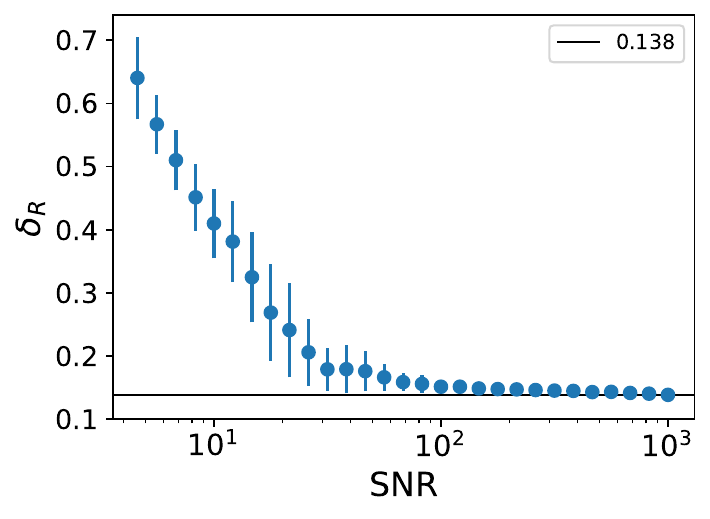}
    \caption{The level of super-resolution achieved by the CFCNN depending on the signal-to-noise ratio using a simulated dataset. The resolvable separation $\delta_\textit{R}$ between two identical emitters is expressed in terms of the Rayleigh distance. A significant spatial resolution enhancement relative to the Rayleigh criterion is achieved across the entire SNR range, exceeding a sevenfold improvement for higher SNR values and attained solely from a single camera image.}
    \label{fig:super-res}
\end{figure}

Furthermore, we generated a standalone dataset consisting of a pair of emitters with progressively increasing separation to assess the level of super-resolution achievable by our CFCNN. The followed procedure is outlined in the Methods section, part E. For the minimum resolvable separation between two emitters reconstructed by the CFCNN as a function of SNR, see Fig.~\ref{fig:super-res}. This separation $\delta_R$ is expressed in terms of the Rayleigh distance. It is evident that a significant level of super-resolution is attained even at the lowest SNR values. For higher SNR values, the achieved spatial resolution saturates towards $\delta_R = 0.138$, corresponding to more than sevenfold enhancement. \rev{Initially, the resolution is limited by the signal-to-noise ratio (SNR), as expected from Cramér–Rao lower bound (CRLB) considerations. As the SNR increases, we observe a saturation behavior imposed by the finite pixel size. The limitations arising from finite pixel size are discussed in detail in the Methods, subsection B. The main result of this analysis is that the $4 \times$ upsampling does not limit the resolution for typical SNRs in fluorescence and single-molecule microscopy applications. A higher upsampling would not yield higher final resolution, and would be costly in terms of computational resources. Let us remark that the conventional resolution limit in single-molecule localization microscopy (SMLM) is derived from the CRLB of individual localizations of two temporally separated (blinking or photoactivated) emitters \cite{Mortensen2010,Thompson2002} and cannot be directly applied in this case. Here, the sources emit simultaneously, and it is known that the Fisher information of their separation vanishes as the emitters get closer~\cite{Tsang2016}, effectively limiting the achievable resolution further.}

For the experimental images of terrylene molecules with an emission peak at $580$~nm, taken with a microscope objective with numerical aperture $1.4$, the CFCNN reaches the resolution of 35~nm.
Furthermore, while SMLM typically achieves up to 10-fold resolution improvement, it requires a long sequence of sparse images and meticulous calibration. By applying the CFCNN to each frame, we could reach the same level of super-resolution while significantly reducing the number of images needed, making the process considerably faster.

\begin{figure}[h]              
    \centering
    \includegraphics[width=0.99\columnwidth]{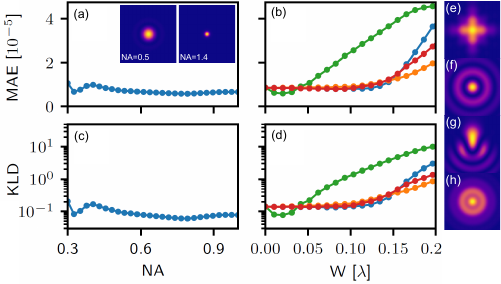}
    \caption{\rev{Reconstruction performance of the CFCNN for varying numerical aperture values in terms of mean absolute error (a) and Kullback–Leibler divergence (c). The results demonstrate that the model maintains a consistently high reconstruction quality across the whole range. \revv{The inset shows ideal PSFs for NA = 0.5 and NA = 1.4.} MAE (b) and KLD (d) between the CFCNN outputs of aberrated star pattern (same as in Fig.~\ref{fig:synth_data}) and the ground truth as a function of the aberration strength $W$ show robustness to moderate optical aberrations. Astigmatism (orange), spherical aberration (red), coma (green), and defocus (blue) were studied individually. \revv{Examples of PSFs distorted by these aberrations are depicted in panels (e--h), respectively, with $W = 0.2~\lambda$.}}}
    \label{fig:aberr}
\end{figure}

\begin{figure*}              
    \centering
    \includegraphics[width=0.99\textwidth]{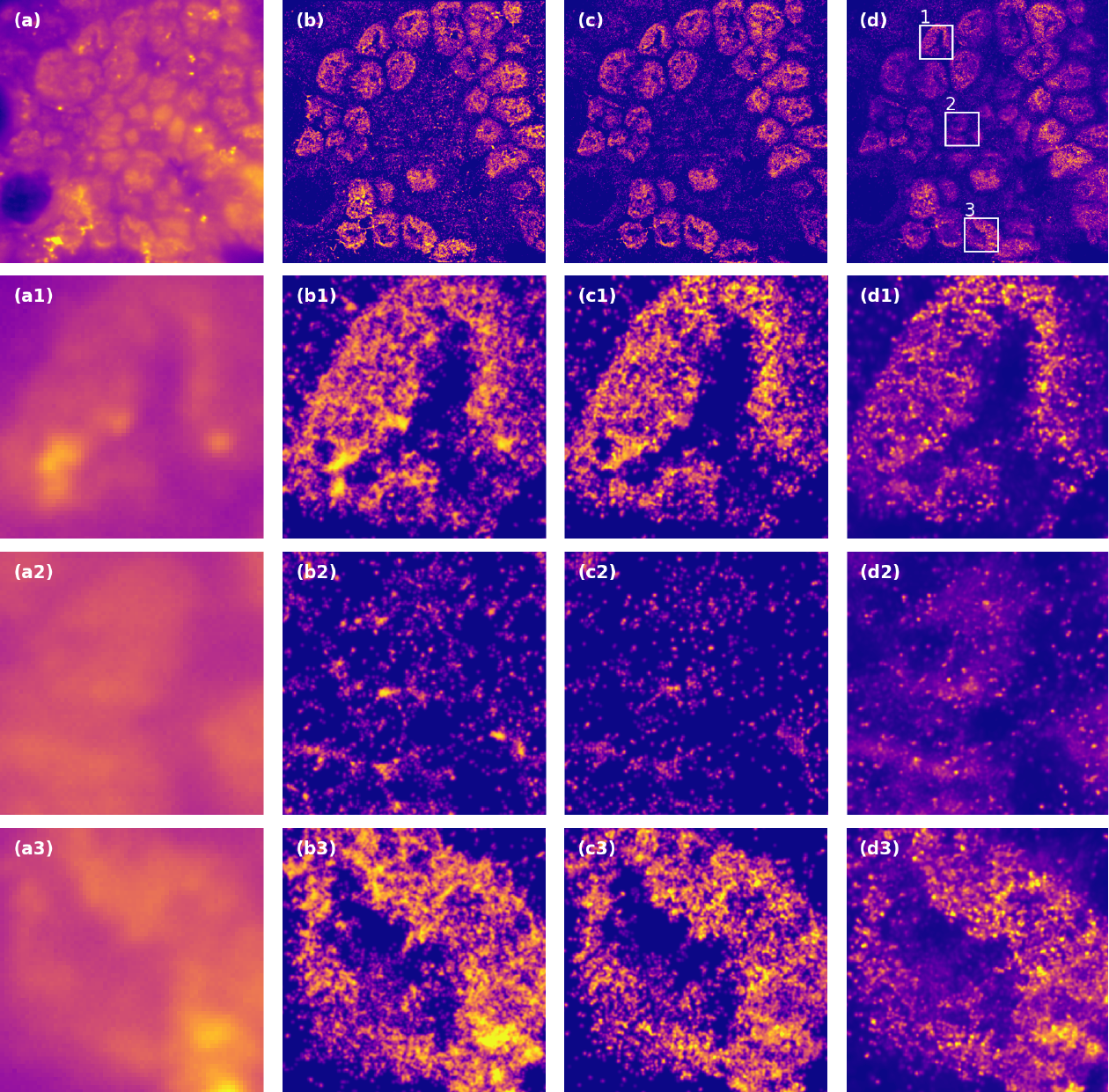}
    \caption{\rev{Demonstration of the CFCNN performance on a real, complex biological data, particularly an image of H4Ac labeled with Alexa Fluor 647 in a mouse intestinal tissue section \cite{Windstorm2019}. \revv{(a) The intensity sum across all raw frames. (b) A resulting image obtained by ThunderSTORM. (c) Result produced by DECODE. (d) CFCNN reconstruction. Three magnified areas are shown in panels (a1--d1), (a2--d2), and (a3--d3).} \revv{The CFCNN yields the most consistent image reconstruction with structures resolved even in the low-SNR central region, where the other methods underperform.}}}
    \label{fig:bio}
\end{figure*}

\rev{So far, we have applied the CFCNN on samples with various SNR values and concentrations of molecules obtained from a high-NA microscope with a near-Gaussian PSF shape. To investigate the independence of the CFCNN on the PSF shape, we performed an additional analysis of low-NA microscopes. The corresponding PSFs were calculated using full vectorial electromagnetic simulations. In Fig.~\ref{fig:aberr}~(a),(c), we show that the performance of our model is essentially independent of the objective numerical aperture, handling the transition from an Airy-type to a Gaussian PSF with the same level of accuracy. This is particularly striking because the CFCNN model was not trained on these intermediate NA values. Together with the experimental results, this analysis clearly demonstrates that our CFCNN maintains its performance across a broad range of microscope configurations and PSF shapes, showing the unprecedented degree of generalization.}

\rev{Since optical setups are often affected by aberrations, we tested our CFCNN (trained only on ideal Gaussian and Airy PSFs) on aberrated PSFs to assess its robustness against such imperfections. In Fig.~\ref{fig:aberr}~(b), (d), we present the image reconstruction quality in terms of MAE and KLD for increasing strength $W$ of aberrations distorting the ideal PSF. Although the CFCNN has never seen any aberrated PSF during training, it can handle a certain level of aberrations with only a slight deterioration in the metric values. Up to $0.15 \lambda$, the MAE and KLD stay virtually constant for spherical aberration, defocus, and astigmatism, while increasing more for coma. \revv{To highlight this visually, PSFs with the studied aberrations of $W=0.2~\lambda$ are shown in Fig.~\ref{fig:aberr} (e)--(h).} These results again underscore the generalization ability of the CFCNN to maintain high performance even under conditions for which it was never explicitly trained, demonstrating resilience to realistic optical imperfections.}
\rev{This resilience can be increased even further by expanding the CFCNN training dataset to aberrated PSFs and, potentially, to other imperfections at the expense of increased computational cost. Finally, it is important to stress that the CFCNN training is performed only once, and the trained model is then readily applicable to any samples and microscope setups without retraining, which represents a completely novel approach to deep-learning super-resolution and significant simplification of any application workflow.}

\rev{To provide another example of the CFCNN performance, and to further substantiate our claims regarding its calibration-free operation and generalization ability to previously unseen data, we demonstrate its application on a real, complex biological sample, specifically H4Ac labeled with Alexa Fluor 647 in a mouse intestinal tissue section with densely packed chromatin structure \cite{Windstorm2019}, with highly heterogeneous background, and unavoidable optical aberrations, see Fig.~\ref{fig:bio}. As in all cases presented throughout this work, the CFCNN did not have access to any information about the data origin. The result from the CFCNN is shown in Fig.~\ref{fig:bio} (d) and compared with ThunderSTORM (b) and DECODE (c).
\revv{Apparently, the CFCNN yields the most consistent image reconstruction with structures resolved across the whole field of view, even in the low-SNR central section, where it identifies many more emitters than the other methods.}}
\revv{The data are visualized after drift correction. In general, the drift correction can be readily applied to the CFCNN reconstructions if necessary.}

\section{Conclusion}

We presented a calibration-free convolutional neural network for single-frame super-resolution fluorescence microscopy. Unlike conventional methods, the CFCNN operates without any calibration, retraining, or prior information about the imaging system, making it highly versatile and universally applicable. Trained on a vast synthetic dataset, the model has the ability to generalize across different experimental conditions, including variations in noise levels, point spread functions, and optical configurations.  We demonstrated its effectiveness on experimental fluorescence microscopy images of terrylene molecules. We compared the super-resolved image reconstructions with the Richardson-Lucy algorithm provided with a precise PSF, the ThunderSTORM multi-emitter fitting with numerous tuned input parameters and specification of the camera properties, \rev{and deep-learning DECODE trained on calibration data corresponding to the optical system and supplied with detailed information about the sample.} We showed that the calibration-free CFCNN consistently outperforms the other methods, on both the experimental images and synthetic data, across all SNR values. The graphical visualization of the super-resolved images clearly demonstrates the superiority of the CFCNN, producing clean and credible reconstructions, while the other methods introduce artifacts and information loss. This visual advantage is supported by quantitative evaluation. Namely, the mean absolute error and Kullback–Leibler divergence consistently demonstrate the superior performance of the CFCNN compared to the system-specific Richardson-Lucy algorithm, multi-emitter fitting, and DECODE methods.

Furthermore, we quantified the super-resolving ability of the CFCNN, achieving up to sevenfold enhancement relative to the standard Rayleigh limit. In our experiment with terrylene, with an emission wavelength of $580$~nm, this corresponds to resolving fluorophores separated by $35$~nm using a single camera frame with an exposure time of 150~ms. The single-frame operation reduces measurement times to dozens of milliseconds, allowing real-time super-resolution imaging of dynamic and live samples that would be inaccessible with slower multi-frame techniques, and photosensitive specimens. Moreover, it relaxes the sparsity condition of blinking emitters, significantly broadening its applicability to a wider range of fluorescent markers. When combined with SMLM approaches, our CFCNN could notably shorten the overall acquisition time by substantially reducing the number of frames, while simultaneously improving accuracy. \rev{We also showed that the CFCNN remains robust against moderate optical aberrations, despite being trained exclusively on ideal PSF shapes, highlighting the strong generalizability of the method.}

The presented work demonstrates the potential of a calibration-free deep learning approach for rapid super-resolution fluorescence microscopy. Future improvements, such as incorporating additional upsampling layers for even higher resolution, introducing residual connections for enhancing the propagation of information \cite{Kaiming}, and using deeper network architectures, could further enhance the versatility and performance of our approach. With these advancements, we envision our network becoming a completely universal tool for high-precision microscopy, broadly applicable across numerous domains ranging from biomedical imaging to material sciences.

\section{Methods}

\subsection{Experiment and sample preparation}

\begin{figure}[h!]              
    \centering
    \includegraphics[width=0.99\columnwidth]{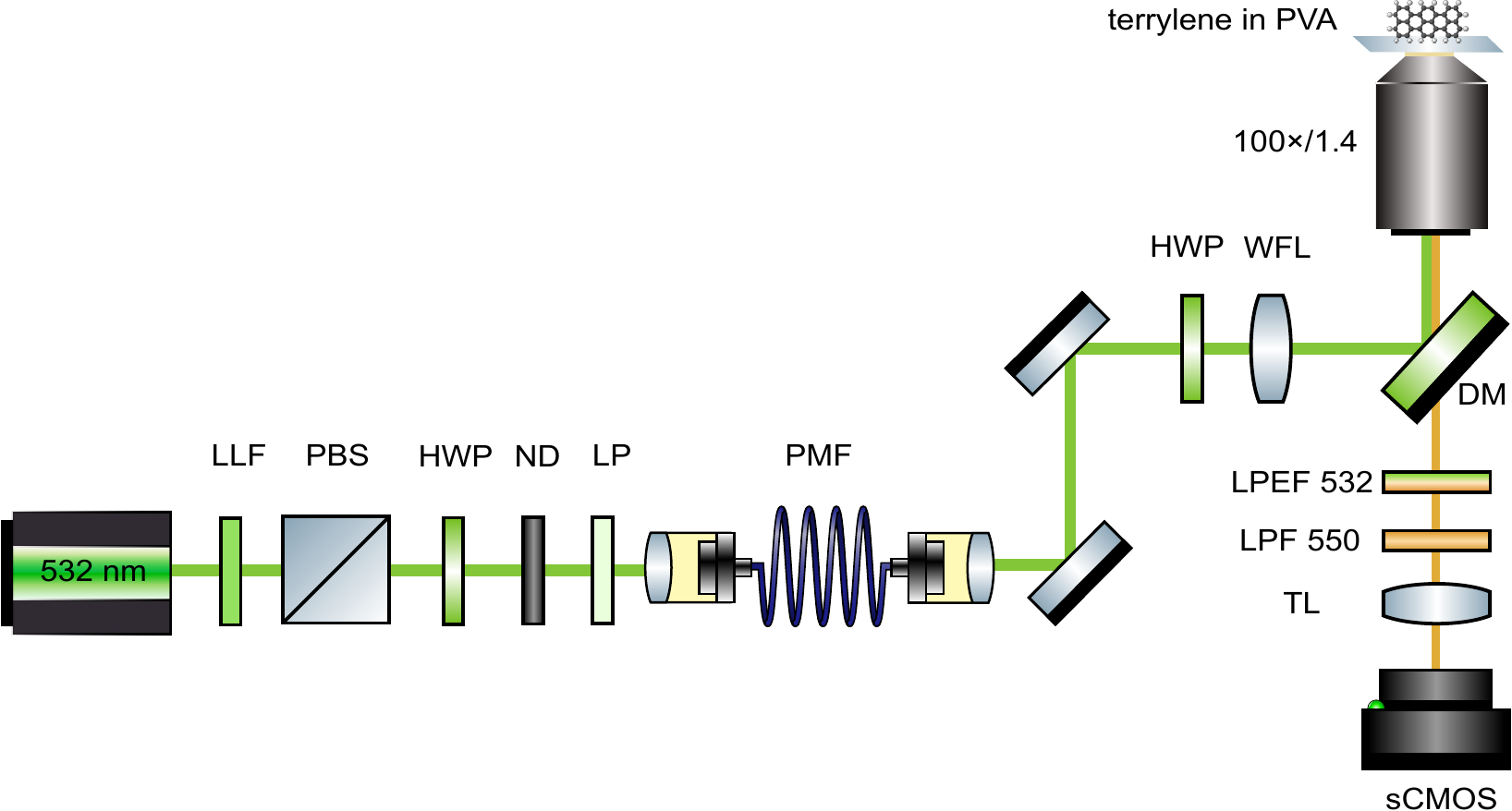}
    \caption{\rev{The experimental setup of a fluorescence microscope. The continuous-wave excitation laser (532~nm) goes through a laser-line filter (LLF) to ensure spectrally pure excitation and its power is set via a polarization control consisting of a half wave plate (HWP) and linear polarizer (LP) and additional neutral density filter (ND) of various optical densities. The excitation path contains a wide-field lens (WFL) for broad sample illumination. The terrylene molecules in the sample are excited through a high-NA objective (100$\times$/1.4). The emitted fluorescence is collected through the same objective and transmitted by the dichroic mirror (DM) to the collection path. A long-pass edge filter (532~nm) and a long-pass filter (550~nm) suppress any residual laser light and the fluorescence is focused by a tube lens (TL, 200~mm) onto an sCMOS camera.}}
    \label{fig:expdet}
\end{figure}

The primary imaging setup consists of a fluorescence microscope operating in a wide-field imaging configuration, equipped with an sCMOS camera (Andor Marana 4.2B-6) with $6.5~\mu$m pixel size. This system enables near-diffraction-limited image acquisition of fluorescent samples. 
\rev{The detailed scheme of the experiment is provided in Fig.~\ref{fig:expdet}.} A $532~$nm excitation laser is directed towards the sample through a high numerical aperture oil-immersion objective lens (OLYMPUS 100×/1.4), allowing optimal excitation and collection of the emitted fluorescence. The Stokes-shifted fluorescence is separated from the excitation light path on a dichroic mirror and detected by the camera, onto which it is focused using an achromatic doublet with a focal length of 200~mm as a tube lens. In combination with the objective of 100× magnification, this results in the overall system magnification of $M \approx 111$.
%\rob{
A laser-line filter is used in the excitation path to suppress unwanted red light from the diode-pumped solid state excitation laser. Two long-pass filters in the collection path prevent any residual laser light from reaching the camera sensor.
%}

The samples contain terrylene molecules embedded in polyvinyl alcohol (PVA) and are prepared by spin coating. The PVA powder was dissolved in 4 ml of warm deionized water to form a 5\%~w/v solution. We added 100~$\textrm{$\mu$}$L of $2 \times 10^{-5}$ \%~w/v terrylene-in-toluene solution and mixed the resulting solution by sonication. Then we spin-coated (30~s at 1500~rpm) a 30-$\mu$L drop of the mixture on a clean coverslip. This technique creates a thin layer of PVA containing the fluorescent molecules across the coverslip. The emission spectrum of terrylene has a peak at $580$~nm.

The PVA forms an environment where terrylene exhibits very limited photostability, with some of the individual molecules undergoing a few blinking cycles before irreversibly photobleaching. We captured a sequence of images from the same location within the sample, with progressively fewer active emitters. The acquisition time of each frame was $150$~ms. Only the first, densest frame was used as input for all the studied methods--our CFCNN, R-L deconvolution, MEF, and DECODE--while the subsequent frames served solely as reference data and were utilized to generate the ground truth, providing a high-resolution benchmark against which the super-resolved outputs of all four methods were compared.

\subsection{Calibration-free convolutional neural network}
\label{sec:meth:cnn}
\begin{figure*}              
    \centering
    \includegraphics[width=0.99\textwidth]{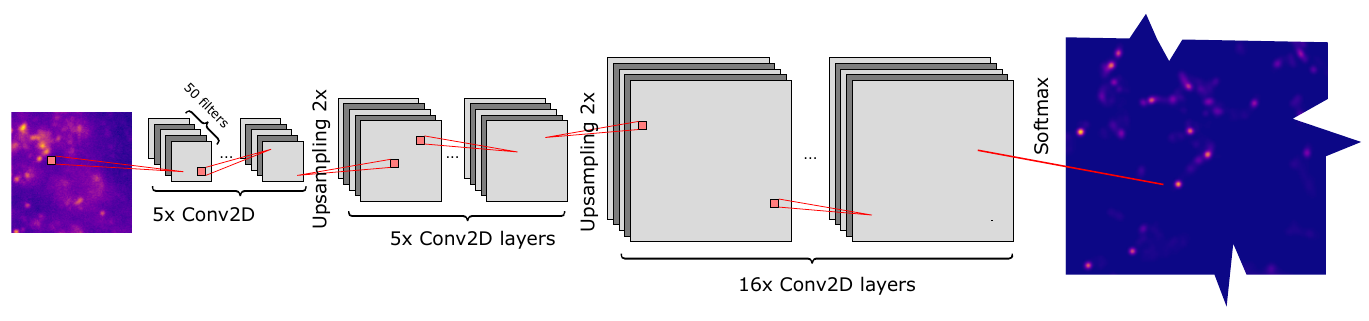}
    \caption{\rev{A visual representation of the CFCNN architecture. The model is fully convolutional, consisting solely of convolutional layers without any fully-connected components. It comprises three convolutional blocks, each containing 50 filters per layer. The blocks are separated by $2\times2$ upsampling layers, which gradually increase the spatial dimensions, resulting in a fourfold magnification of the input image. The first two blocks each contain five convolutional layers, while the final block includes 16 layers for more precise reconstruction at the finer scale. See Methods, section B, for more details.}}
    \label{fig:cnnarch}
\end{figure*}

We designed a calibration-free deep learning model for super-resolution imaging. This fully convolutional neural network reconstructs a high-resolution image from a single intensity frame. \rev{The CFCNN architecture is illustrated in Fig.~\ref{fig:cnnarch}}. Consisting of 26 hidden convolutional layers with 50 filters each, the model performs a nonlinear transformation on the input. Information propagates through $5 \times 5$ convolutional kernels, performing localized feature extraction across layers. To implement nonlinearity and prevent the dying ReLU problem \cite{ReLU}, we employ a LeakyReLU activation function with a negative slope of 0.3. The final layer uses softmax activation \cite{softmax} to enforce the normalization and non-negativity of the output. Each convolutional layer includes dropout regularization, randomly setting 1\% of input unit values to zero during training to mitigate overfitting \cite{dropout}. Additionally, we integrate upsampling layers to incorporate interpolative resizing into the network architecture. Positioned after the 5th and 10th convolutional layer, each upsampling doubles the input dimensions, resulting in a fourfold image magnification. Furthermore, due to the fully convolutional-based design, the model can process images of arbitrary shapes and sizes, making it highly adaptable for various imaging applications.

\rev{The fixed $4\times$ upsampling factor may restrict the achievable localization accuracy and the effective optical resolution. However, both are fundamentally limited by the optical system, noise statistics, and available signal strength, as described by the Cramér–Rao lower bound (CRLB) \cite{Thompson2002}. For temporally separated (blinking or photoactivated) emitters in SMLM datasets, the attainable optical resolution is approximately 2.3 times the localization precision \cite{Mortensen2010}. Ideally, the upsampling factor should be chosen according to the CRLB for the given experimental conditions, as finer grids increase computational cost without necessarily improving resolution. Based on the typical emitter brightness and noise levels in our data, we estimate a theoretical localization precision of approximately 6 nm, corresponding to half of the upsampled pixel size in object space. The resulting localization-based optical resolution is therefore limited to about 14 nm, consistent with the effective upsampled pixel spacing. Since the CRLB represents an optimistic lower bound \cite{Mortensen2010}, this choice reflects a reasonable trade-off between achievable resolution and computational efficiency.}
\rev{Surprisingly, numerical simulations further indicate that the localization accuracy is not strictly limited to one quarter of the original pixel size, but can reach approximately one tenth. This improvement arises from the CFCNN’s ability to interpolate: when an emitter lies between two adjacent upsampled pixels, the network assigns non-negligible intensity to both, and in the high-SNR limit, the centroid of this distribution yields a more precise position estimate. In our system, this corresponds to a theoretical localization accuracy of 6 nm, which directly translates to localization-based resolution. Finally, for SNR levels in our data, which are typical for fluorescence and SMLM microscopy, the resolution is not limited by the finite ($4 \times$) upsampling of the CFCNN.}

We train the model using a simulated dataset of low-resolution $50 \times 50$ pixel images paired with their corresponding $200 \times 200$ high-resolution ground truth representations. The dataset is generated using diverse optical conditions to ensure broad applicability across different imaging scenarios. To create each sample, we randomly assign emitter intensities and positions within a $200 \times 200$ pixel grid. The number of emitters in a sample is randomly selected from the $\left[ 1, 25 \right]$ range, while their intensities follow a Poisson distribution with a mean value ranging between $\left[ 1, 3000 \right]$ photons. The resulting matrix represents our ground truth image.
Next, we simulate the effects of finite resolution by convolving the ground truth matrix with a point spread function. We consider two distinct PSF shapes: A Gaussian function, commonly used to approximate high numerical aperture imaging conditions, and an Airy disc, representing lower numerical aperture scenarios \cite{GaussPSF}. The full width at half maximum of the PSF varies between approximately $\left[ 2.5, 7.5 \right]$ pixels in the $50 \times 50$ grid, equivalent to $\left[ 10, 30 \right]$ pixels in the $200 \times 200$ grid. Moreover, we introduce PSF asymmetry by applying a squeezing transformation along a random axis.
Following the convolution, the image is downsampled to $50 \times 50$ pixels, and background noise is added, with intensity values randomly selected from the $\left[ 1, 100 \right]$ photon range per pixel. Each dataset sample is generated independently, ensuring a unique combination of the number of emitters, intensity, noise level, PSF shape, width, and asymmetry.

To train the model, we employed incremental learning, a dynamic approach where newly generated data pairs gradually replace the existing samples during training \cite{inclear}. This combination of incremental learning and data simulation is particularly beneficial for developing universal networks capable of handling diverse imaging conditions. Throughout 90 incremental iterations, the model processed nearly 500 000 data pairs, with each iteration consisting of 50 epochs of training on mini-batches of 32 samples. The training was performed using the Adam optimizer, which leverages adaptive moment estimation for efficient gradient descent \cite{Adam}. The initial learning rate was set to $10^{-4}$. Given the high sparsity of the ground truth images, we used an adjusted mean squared error loss function. First, both the ground truth $I$ and the model prediction $\hat{I}$ are convolved with a Gaussian filter $g$ of $\sigma = 2$. Then, we compute the squared difference between the filtered images and average it over a batch of $N$ samples, $\frac{1}{N} \sum_{i=1}^{N} \Vert I_i \ast g - \hat{I}_i  \ast g \Vert_{2}^{2}$. Similarly, an adjusted mean absolute error metric, $\frac{1}{N} \sum_{i=1}^{N} \Vert I_i \ast g - \hat{I}_i  \ast g \Vert_{1}^{2}$, was used to monitor convergence and evaluate the model performance. The presented hyperparameter configuration results from manual optimization of network architecture and dataset ranges to enhance the performance across the device-independent regime. We expect further improvement by using a larger network architecture. However, the computational resources, namely the required operating memory for its training, are presently beyond our technical capacities. The CFCNN model, code, and data are available at the public repository \cite{zenodo}.

\subsection{Richardson-Lucy algorithm, ThunderSTORM, \rev{and DECODE}}

We employed the Richardson-Lucy algorithm \cite{R-L_R, R-L_L}, multi-emitter fitting using ThunderSTORM \cite{ThunderSTORM}, \rev{and DECODE \cite{DECODE}} on both the experimental fluorescence microscopy images and synthetic data as relevant representatives of state-of-the-art methods for comparison with our deep learning approach.

Richardson-Lucy algorithm recovers the true image by performing deconvolution, that is, reversing the blurring caused by a known point-spread function. It is an iterative procedure, maximizing the likelihood that the observed blurred image corresponds to the true image convolved with the PSF. Therefore, the precise PSF, an initial guess, and the number of iterations or error threshold must be specified and are user-dependent. In our application, a uniform image was chosen as a starting point, and the iterations were stopped when the MAE between consecutive iterations satisfied $\epsilon \leq 10^{-10}$. To achieve the $200 \times 200$ pixel output size, we upscaled the camera image using bicubic interpolation prior to applying the R-L reconstruction. Upscaling before reconstruction ensures that the R-L algorithm operates on the image at the target resolution, preserving the intended precision. Performing the reconstruction first and interpolating the result afterward would handicap the method by introducing artifacts, which could distort the fine details in the deconvolved image. For each analyzed dataset of blinking emitters, only the single densest image was given to the reconstruction algorithm. The rest of the sequence provided a reference where a few isolated emitters were fitted with a Gaussian PSF to set the optimal kernel for the deconvolution.

ThunderSTORM is an open-source ImageJ plugin designed to analyze data obtained through SMLM techniques. The default settings of the software are to produce reasonable results on various datasets. To optimize the results, the users can adapt the analysis to suit their particular data using the included processing and post-processing methods and tools. This customization, however, requires experience and extensive knowledge about the system under study. While users usually have this knowledge, the process is time-consuming and complex, involving trade-offs that must be reconsidered whenever conditions change. Moreover, the same settings may not be optimal across the entire field of view and should always be adjusted for different experimental realizations to yield the best results. In our analysis, we enabled multi-emitter fitting and specified the details of our imaging setup and camera (photoelectrons per A/D count, etc.). When applied, the method returns a table as a result, which contains localized emitter coordinates and intensities. We generated the image by rendering the information from the table into a $200 \times 200$ pixel grid using the nearest pixel method.

\rev{DECODE is a deep-learning framework for super-resolution fluorescence microscopy that reconstructs emitter positions and intensities by leveraging a convolutional neural network. The utilization of the method is not straightforward, as the user must complete several demanding steps to obtain a model that is not universal, and retraining is necessary whenever experimental conditions change. The user trains DECODE for their particular application and setup, which requires a lot of input information. Specifically, a calibration PSF $z$-stack needs to be provided to the external software SMAP, which fits the PSFs, and the fitted parameters are used by DECODE in the training. Therefore, the $z$-stack of a specific PSF must be measured or generated. In addition, the training procedure requires specifying numerous parameters, including emitter density, brightness, camera characteristics, spatial ($x$, $y$, $z$) and intensity ranges, as well as the level of fluorescent background. The training itself typically takes several hours, and the resulting model can only be applied to images with parameters that match those used during training, in all size, shape, and intensity. We trained DECODE separately for our various datasets. In case of the experimental data, we generated a fine-tuned PSF accurately matching the experimentally measured PSFs, and produced the $z$-stack by applying defocus via Zernike polynomials. This was used for the training, along with all specific information about our optical setup.}
\rev{For the synthetic data (Fig.~\ref{fig:synth_data} of the main text), we trained DECODE on the specific PSF and SNR values separately for each red data point in Fig.~\ref{fig:synth_data} (m, n), so that it had the best knowledge of all the parameters possible.}

\rev{We found that each trained DECODE model introduces a systematic shift in the pixel positions of the localized emitters, which is consistent across all reconstructions performed by the same model. We corrected for this shift when evaluating the reconstruction quality metrics. Even after applying this correction (which differs between models and is therefore inherently introduced during DECODE’s training), it is evident in the Results chapter that our CFCNN achieves superior performance in all cases.
Furthermore, the training of the DECODE model is not fully reproducible. When initiated with the same calibration data and the same set of parameters, repeatedly trained DECODE models exhibit varying performance on a test dataset, even though all models pass the DECODE built-in convergence check functionality. In light of this fact, our calibration-free super-resolution approach provides a significant benefit to practitioners in biomedical imaging. Instead of a time-consuming and challenging training process, a user is supplied with an already trained final model, which can be directly applied to a wide variety of microscope setups and samples.}

It is important to note again that all the studied methods were provided only with a single densest fluorescence image of organic molecules shown in Fig.~\ref{fig:tiPVA}~(a). We recognize that both ThunderSTORM \rev{and DECODE}, as well as our CFCNN, would achieve notably better performance with access to the complete sparse image sequence. However, our objective is to compare these methods on dense samples without relying on additional data or imposing specific experimental requirements.

\subsection{Ground truth determination}

\begin{figure}[ht]              
    \centering
    \includegraphics[width=0.99\columnwidth]{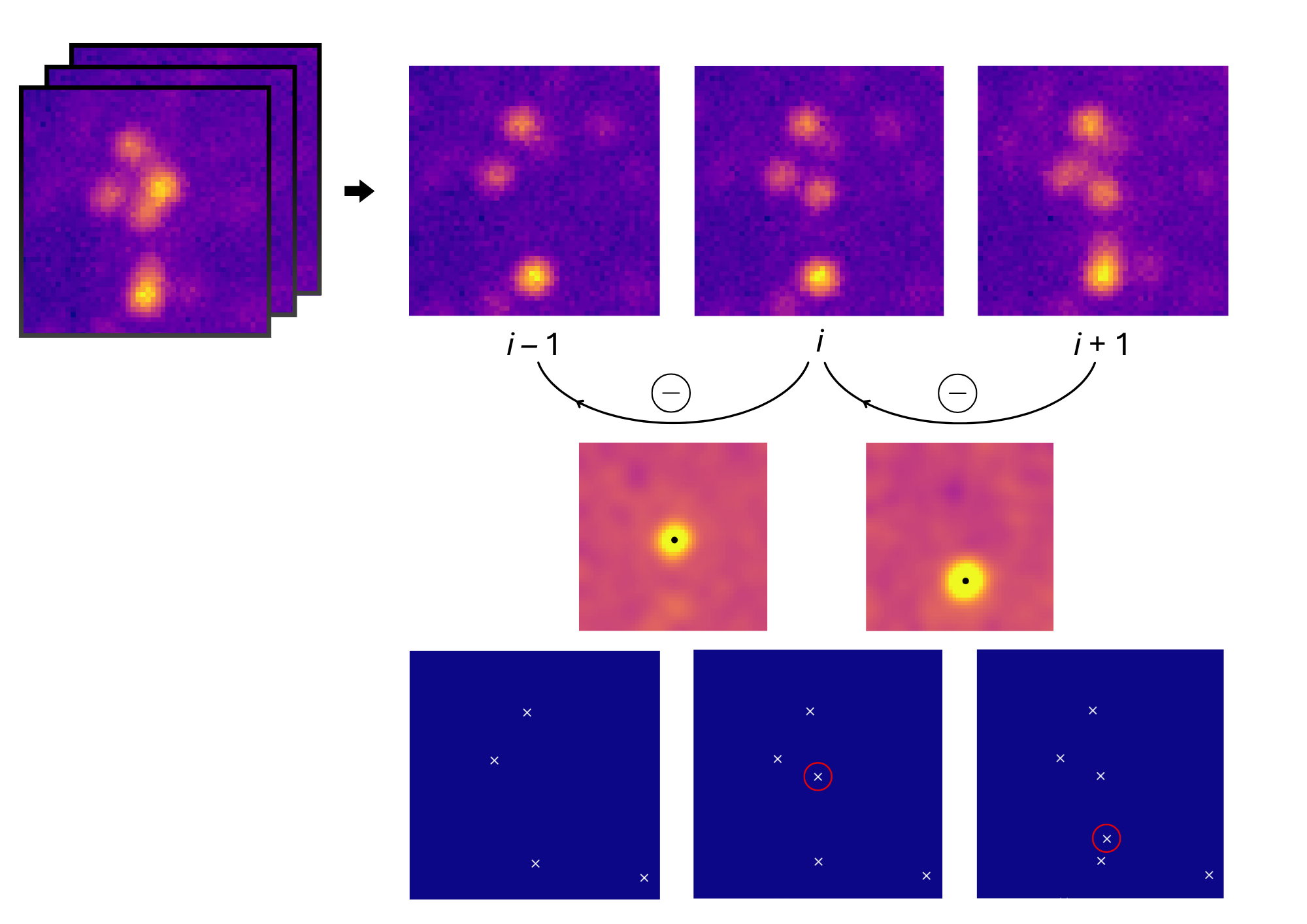}
    \caption{\rev{Schematic illustration of the ground truth determination workflow. The terrylene molecules are individually localized from differential images of a reversed bleaching sequence, see Methods, section D, for details.}}
    \label{fig:gtdet}
\end{figure}

When testing a super-resolution method on experimental data, the fundamental challenge is the lack of an absolute ground truth. Unlike simulated datasets, where ground truth information is inherently available, real fluorescence microscopy data provide no direct reference for evaluating reconstruction accuracy. To address this, we leveraged the limited photostability of terrylene in PVA to evaluate the performance of all methods on the realistic fluorescence microscopy dataset. \rev{The workflow is illustrated in Fig.~\ref{fig:gtdet}.} Sequential images of the sample were captured, with emitters gradually photobleaching over time. This natural reduction in emitter density allowed the application of a standard localization method on the emitters that became isolated and distinguishable throughout the sequence and on the temporal difference of the frames. 
For the ground truth construction, we began with the final image of the sequence, where only a few distinguishable emitters remained active. These emitters were fitted with Gaussian profiles, enabling precise localization of their positions and intensities. This final frame was subtracted from the preceding one to produce another sparse image, allowing additional emitters to be isolated and localized. This process was iterated over the image sequence, progressively building a localization table, which was then rendered into a $200 \times 200$ pixel ground truth image using the nearest pixel method. By starting from the sparsest frame and working backward, this approach minimizes localization errors and ensures consistency across the sequence. 

The resulting manual localization serves as a benchmark for evaluating the performance of the studied methods on the realistic fluorescence microscopy dataset. For quantitative comparison, we calculated two metrics between the visualized images $A = A' \ast g$ and $B =  B' \ast g$, where $A', B'$ denote the reconstructed images and the ground truth, respectively, which are convolved with a narrow Gaussian visualization filter $g$ of $\sigma = 2$. All the presented images are normalized to a unit $L_1$ norm after subtracting the minimum intensity so that the relative intensity distribution is retained,
\begin{equation}
    P_i(A) = \frac{A_i - \min{A}}{\sum_{j=1}^{N} \left( A_j - \min{A} \right) },
\end{equation}
where $N$ is the number of pixels. The mean absolute error (MAE) 
\begin{equation}
    \textrm{MAE} = \frac{1}{N} \sum_{i=1}^{N} | P_i(A) - P_i(B) |
\end{equation}
quantifies the average absolute difference between corresponding pixel values of two images.
The Kullback-Leibler divergence (KLD) 
\begin{equation}
    \textrm{KLD} = \sum_{i=1}^{N} P_i(B) \log_2 \left( \frac{P_i(B)}{P_i(A)} \right)
\end{equation}
evaluates the difference between the probability distributions of the pixel values in the two images. To estimate statistical errors, we used the following approach. Our bootstrapping model consists of Poissonian emitters on top of a normally distributed background. We estimated the parameters of background distribution from data regions without emitters, and we determined the $\lambda$ value of the Poissonian signal as a distance from the mean value of the background. This model accurately approximates the statistical noise characteristics of our camera. We generated 100 random samples with that model and inserted them into tested reconstruction methods to estimate the standard deviation for each tested metric. Such an error bar mainly determines the numerical stability of the reconstruction method.

\subsection{Synthetic data and super-resolution}

To systematically evaluate and compare the performance of the studied methods, we generated synthetic datasets designed to mimic a wider range of experimental conditions, for example, the background noise level. These datasets consist of simulated fluorescence images with varying emitter numbers, intensities, spatial distributions, structural complexity, and noise levels.

The emitters were positioned in specified patterns within a defined field of view, serving as the absolute ground truth. The availability of this exact ground truth is a key advantage of the simulated datasets. The synthetic data thus provides a fully reliable reference for method evaluation.
To simulate the optical system, we convolved the $\delta$-functions at the emitter positions with a PSF modeled as a two-dimensional Gaussian function with a standard deviation corresponding to the imaging conditions, replicating the diffraction-limited appearance of molecules in fluorescence microscopy. 
To assess robustness against experimental imperfections, we incorporated different levels of Gaussian background noise. The SNR was systematically varied to study its impact on reconstruction quality.

Using the synthetic data, we demonstrated the performance of our calibration-free convolutional neural network, R-L deconvolution, and MEF using ThunderSTORM, depending on the amount of background noise in the input image by varying the SNR. The reconstruction accuracy was again quantified by computing MAE and KLD between the output images and the ground truth. Monte Carlo simulations were performed to estimate error bars for each data point.

Furthermore, we assessed the super-resolution achievable by our CFCNN. To this end, we generated a dataset consisting of emitter pairs with identical intensity. In the first frame, the emitters were perfectly overlapping, and their separation was incrementally increased in subsequent frames within the range of $\left[0, 1\right]$ of the Rayleigh distance, with a step size of $1/80$ of the Rayleigh distance. Since the Rayleigh criterion is based on the first minimum of an Airy pattern, this distance is approximately equivalent to three standard deviations of Gaussian emitter profiles. 
Each frame in the sequence was processed by the CFCNN, and the resulting images were analyzed by their marginal intensity distributions. The modulation in these distributions determines the CFCNN’s ability to resolve closely spaced emitters. Following the Rayleigh criterion, the emitters are resolved if the modulation dip has a relative depth of 26.3\%. The same sequence was analyzed across an SNR range of $\left[1, 1000\right]$ with a logarithmic step and 100 Monte Carlo samples per data point.

\rev{\subsection{Optical aberrations}}
\rev{Optical systems are inevitably affected by aberrations, which can degrade image quality and challenge reconstruction accuracy in fluorescence microscopy. To assess the robustness of our CFCNN under such conditions, we evaluated its performance on images generated with various aberrated PSFs. Importantly, the network was trained solely on ideal Gaussian and Airy PSFs, without exposure to any aberrated data. This analysis allows us to determine how well the CFCNN generalizes to realistic optical imperfections and maintains super-resolution reconstruction quality despite deviations from the ideal imaging conditions.}

\rev{To systematically investigate the impact of optical aberrations on the CFCNN performance, we considered four common types: astigmatism, coma, spherical aberration, and defocus. These aberrations were modeled within a vectorial PSF framework \cite{Richards1959}, with Zernike polynomials \cite{Gu1999} employed to parameterize and control the aberration strength in units of $\lambda$. We utilized the synthetic star dataset, applied each aberration individually, and evaluated the MAE and KLD between the ground truth and the CFCNN reconstructions of the aberrated images as a function of aberration strength.}

\vspace{10mm}

\section*{Acknowledgments}
We acknowledge the use of cluster computing resources provided by the Department of Optics, Palacký University Olomouc. We thank J. Provazník for maintaining the cluster and providing support.

\section*{Data availability}

The scripts and data analyzed in this study are publicly accessible on Zenodo \cite{zenodo}.

\section*{Funding}
This work was supported by the Ministry of Education, Youth, and Sports of the Czech Republic (project OP JAC CZ.02.01.01/00/23$\_$021/0008790) and the Czech Science Foundation (project 21-18545S). AD and DV acknowledge the support by Palacký University Olomouc (projects IGA-PrF-2024-008, IGA-PrF-2025-010, and IGA-PrF-2026-005).

\section*{Author contributions}
AD and RS performed the experiments and numerical simulations. DV developed the deep learning model. AD drafted the manuscript. MJ conceived the idea of calibration-free super-resolution microscopy and supervised the project. All authors contributed to data interpretation and manuscript revisions.

\section*{Conflict of interest statement}
The authors declare no conflicts of interest.

\bibliography{cfcnn}

\end{document}